\documentclass[aps,pra,twocolumn,showpacs,amsmath,amssymb]{revtex4}

\usepackage{amsmath}
\usepackage{amssymb}
\usepackage{graphicx}
\usepackage{epsf}
\usepackage{flafter}%
\usepackage{bm}

\begin{document}


\title{Long-range potentials and $(n-1)d+ns$ molecular resonances in an ultracold Rydberg gas}

\author{J. Stanojevic}
\author{R. C\^{o}t\'{e}}
\author{D. Tong}
\author{E.E. Eyler}
\author{P.L. Gould}

\affiliation{Department of Physics, University of Connecticut, Storrs, CT 06269, USA}

\begin{abstract}
We have calculated long-range molecular potentials of the $0_g^{+}$, $0_u^{-}$ and $1_u$ symmetries between highly-excited rubidium atoms. Strong $np+np$ potentials characterized by these symmetries are important in describing interaction-induced phenomena in the excitation spectra of high $np$ Rydberg states. Long-range molecular resonances are such phenomena and they were first reported in S.M. Farooqi {\it et al.}, Phys. Rev. Lett. {\bf 91} 183002. One class of these resonances occurs at energies corresponding to excited atom pairs $(n\!-\!1)d+ns$. Such resonances are attributed to $\ell$-mixing due to Rydberg-Rydberg interactions so that otherwise forbidden molecular transitions become allowed. We calculate molecular potentials in Hund's case (c),  use them to find the resonance lineshape and compare to experimental results.
\end{abstract}
\maketitle
\section{\label{sec:intro}Introduction}
%
%
Rydberg atoms have long been studied for their unique properties, such as long radiative lifetimes, large cross sections and huge polarizabilities \cite{gallagher-book}. At high principal quantum numbers, interaction forces between Rydberg atoms become extremely large. Manifestations of these long-range interactions can be seen in ultracold collisions \cite{oliveira03} or density-dependent line broadening of resonances in atomic beams
\cite{raimond81}. In ultracold Rydberg systems thermal motion is greatly reduced so that the effects of strong interactions can be investigated and utilized in a more controlable manner. The prospect of using ultracold Rydberg atoms in quantum information applications has stimulated a great deal of experimental and theoretical interest lately \cite{jaksch00,lukin01}. It  was proposed to use the effect of dipole blockade to realize scalable quantum gates  \cite{lukin01}. In mesoscopic ensembles of atoms, one excited atom prevents the Rydberg excitation of its neighbors because strong interactions shift many-atom excited states out of resonance. Recently, large inhibition of Rydberg excitation due to  van der Waals (vdW) interactions (for $n\!\sim$ 70-80), have been observed using a pulse-amplified single-mode laser \cite{tong04} and cw excitation \cite{singer04,Liebisch}. Also, the dipole-blockade of cw Rydberg excitation of Cs atoms has been achieved \cite{voght06}.    

There have been several proposals for weakly bound states involving Rydberg atoms but they have not yet been detected. The so called ``trilobite" and ``butterfly"  states are molecular states formed by a pair of atoms, with one of the atoms in the ground state and the other one in a Rydberg state \cite{greene00,granger01,hamilton02,chibisov02,khuskivadze02}. Bound states of two Rydberg atoms have been  also proposed \cite{boisseau02}. The long-range  molecular resonances are another effect of Rydberg interactions. In the experiment \cite{farooqi03}, $(n\!-\!1)p+(n\!+\!1)p$ and $(n\!-\!1)d+ns$ resonances have been observed in single photon UV excitation from the $5s$ ground state to high $np$ Rydberg states ($n=50-90$).  A detailed theoretical treatment of $(n\!-\!1)p+(n\!+\!1)p$ resonances was presented in \cite{Stanojevic}. However, the energy separations betwen the $(n\!-\!1)d+ns$  resonances and the atomic $np$ resonances are several times greater than the corresponding energy spacings for the $(n\!-\!1)p+(n\!+\!1)p$ resonances. As a result, much higher laser power is needed to excite the $(n\!-\!1)d+ns$  resonances. Also,  many more molecular states are needed for a detailed analysis
of the interaction-induced  $\ell$-mixing, which is necessary for the existence of the resonances.

\section{\label{sec:theory}Theory}

This paper consists mainly of two parts. In the first part we calculate long-range molecular potentials of the $0_g^{+}$, $0_u^{-}$ and $1_u$ symmetries since these symetries describe strong $np+np$ potentials. In the second part we evaluate the excitation dynamics of these molecular states to investigate the variouseffects of interactions  between the atoms. In this paper we primarily study the $(n-1)d+ns$ resonances reported in \cite{farooqi03}. They occur at the average energy of excited atom pairs $(n-1)d$ and $ns$, and do not correspond to any single-atom transitions. The contributions of diatomic potentials which coincide with the asymptotic $(n-1)d+ns$ levels are dominant, although, there are significant contributions from $(n-2)d+(n+1)s$ and $(n-2)p_{1/2}+(n+2)p_{1/2}$ potentials, with approximately the same asymptotic energies. In one-photon transitions from the 5$s$ ground state, dipole transitions to $nd$ and $ns$ states are not allowed. However, at high principal quantum numbers, long-range Rydberg-Rydberg interactions cause $\ell$-mixing so that other molecular states, besides $np+np$, become accessible. Although the physics of this resonance and the $(n-1)p+(n+1)p$ one is very similar,  the treatment of  $(n-1)d+ns$ resonances is technically much more demanding. There are many more asymptotic states between  $np+np$ and $(n\!-\!1)d+ns$ asymptotes and they all have to be included in order to describe the $\ell$-mixing  correctly.
To make the potentials accurate at short distances,  many nearby asymptotes must also be included in the asymptotic basis. In addition, the laser intensity used to excite them was almost two order of magnitude greater than that used for $(n-1)p+(n+1)p$ resonances so there could be more power-dependent terms  to consider, besides the two-photon Rabi frequency. We have used several different approaches to evaluate the results of the Rydberg excitation of atom pairs and they all give consistent results.

\subsection{Long-Range Rydberg-Rydberg potential curves}

We calculate long-range potentials in Hund's case (c) by diagonalization of an interaction matrix. There are several reasons why this Hund's case is considered. Here we point out  only the basic arguments  since they are explained in detail in \cite{Stanojevic}.  We consider the states that are directly or indirectly coupled to $np+np$ asymptotes, which can be directly excited from the $5s+5s$ ground state. We can focus on strong potentials only because they can significantly mix, at short internuclear separations, with other asymptotic states. Such $np+np$ potentials are those coupled to nearby $ns+(n+1)s$ states. The asymptotic energy spacing between $np+np$ and $ns+(n+1)s$ states is quite small, which additionally increases the effect of their strong dipole-dipole coupling. If there were no atomic fine structure, the only candidates for strong $np+np$ potentials would be the asymptotically degenerate ${}^{1}\Sigma_g^{+}$ and ${}^{3}\Sigma_u^{+}$ states \cite{singer04}. After adding spin, the possible symmetries with strong potentials are  $0_g^{+}$, $0_u^{-}$ and $1_u$. Fine structure can be, at least initially, neglected only if its energy splitting is much less than the separation between adjacent $n\ell + n'\ell'$ asymptotic levels, which is not the case here. For instance, the asymptotic energy spacing between $70p_{1/2}\!+\!70p_{1/2}$ and
$70p_{3/2}\!+\!70p_{3/2}$ is almost three times larger then the asymptotic separation between $70s_{1/2}\!+\!71s_{1/2}$ and $70p_{3/2}\!+\!70p_{3/2}$. This obstacle impedes a unique definition of the dispersion coefficient $C_6$. However, these dispersion coefficients cannot be used because the interaction energy of pairs of Rydberg atoms contributing  to $(n\!-\!1)d+ns$ resonances is about twenty times greater than the region of energy for which perturbation theory is applicable. To obtain potential curves one has to diagonalize simultaneously both interaction terms, the long-range Rydberg term and the atomic fine-structure one. 

The interaction matrix includes long-range Rydberg-Rydberg interactions $V_{\rm Ryd}(R)$ and 
the atomic spin-orbit interaction $H_{\rm fs}$
\begin{equation}\label{Utotal}
 U(R)=V_{\rm Ryd}(R)+H_{\rm fs} \;.
\end{equation}
The eigenproblem of this interaction matrix is solved for the $0_g^{+}$, $0_u^{-}$ and $1_u$ states only. 
The first step is to construct asymptotic states of these symmetries and used them as a basis to represent $U(R)$. The projection $\Omega=m_1+m_2$ of the total angular momentum onto the molecular axis is conserved and properly symmetrized asymptotic states are constructed as follows
\begin{eqnarray}\label{symmetrization}
 | n\ell_j,m_j;n'\ell'_{j'},\Omega-m_j;\Omega_{g/u}\rangle && \nonumber \\
 &&\hspace{-1.85in} 
 \sim \left[ | n,\ell,j,m_j\rangle | n',\ell',j',\Omega-m_j\rangle 
      \rule{0.in}{.15in} \right. \nonumber \\ 
 && \hspace{-1.9in} \left. \rule{0.in}{.15in} 
 -p (-1)^{(\ell+\ell')} | n', \ell',j' ,\Omega-m_j\rangle 
                        |n, \ell, j ,m_j\rangle \right] .
\end{eqnarray}
No overlap of the charge distributions of different atoms is assumed here since ultracold gases are very dilute, even considering these Rydberg atoms. All molecular states are defined in the molecule-fixed reference frame. States with $\Omega\!=\!0$ have to be (anti)symmetric 
under the action of the reflection operator $\sigma_{\nu}$. We have addopted the following convention \cite{Brown} for 
the action of $\sigma_{\nu}$ 
\begin{align}\label{reflection-def}
 & \sigma_{\nu}\left|\Lambda\right> = 
        (-1)^{\Lambda}\left|-\Lambda\right> \; , \\
 & \sigma_{\nu}\left|S,M_S\right>   = 
        (-1)^{S-M_S}\left|S,-M_S\right> \; .
\end{align}
The first rule (\ref{reflection-def}) is obviously not applicable if $\Lambda=0$. The correct result of $\sigma_{\nu}$ on such states follows from its action on atomic states  $\sigma_{\nu}\left|\ell,m\right> = (-1)^{m}\left|\ell,-m\right>$. 
\begin{figure}[t]
   \centerline{\epsfxsize=3.4in\epsfclipon\epsfbox{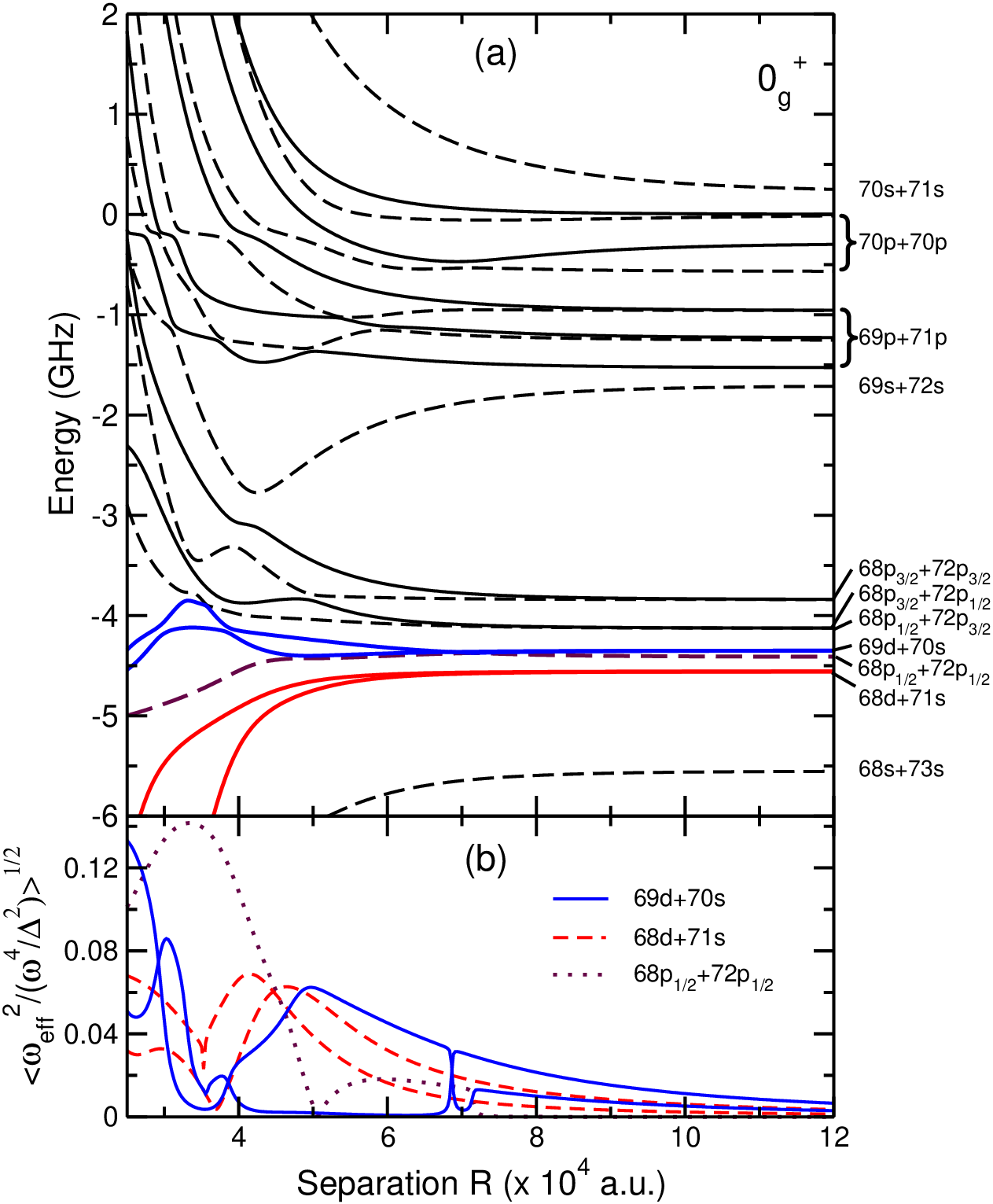}}
\caption{\label{0Gsymm}(a)  Potentials curves for the $0_g^{+}$ symmetry. We present all the potential curves between $70s+71s$ and $68s+73s$. Molecular states, besides those that coincide with the asymptotic $70p+70p$ states, may become accessible due to $\ell$-mixing induced by interactions. The resonance is dominantly produced by the $69d+70s$ states, although significant contributions also come from the  $68d+71s$ and $68p_{1/2}+72p_{1/2}$ states. There are two $69d+70s$ and $68d+71s$  states of this symmetry and only one from the $68p_{1/2}+72p_{1/2}$ asymptote. The fine structure of  $69d$ and $68d$ states is ignored.  (b) Average radial dependencies of the two-photon Rabi frequency. The excitation probability of an atom pair depends on its separation $R$ since the fraction of $p$-character of any molecular state is a function of $R$. The radial dependence is shown for all states of the three asymptotes contributing to the resonance.}
\end{figure}

After adding spin, there are many more different symmetries to consider. On the other hand, it makes the $\ell$-mixing a smoother function of interaction energy, simply because there are many more potential curves with a given $\ell$-component for a given range of energy. We are primarily interested in how $\ell=1$ or $p$-character is distributed over different potential curves for each $R$. If  $\alpha(R;\lambda)$ is the fraction of a given asymptotic state $|\phi\rangle$ in the molecular state $|\varphi_{R;\lambda}\rangle$, then the sum of $\alpha(R;\lambda)$ over all molecular states has to be unity for each $R$. This statement is just the normalization condition of $|\phi\rangle$ in the eigenbasis of $U(R)$. We use this $\alpha(R;\lambda)$ as a measure of the character associated with $|\phi\rangle$. Therefore, for any $np+np$  asymptotic state, the sum of the $p$-character of all molecular states has to be unity for any $R$.
\begin{figure}[b]
   \centerline{\epsfxsize=3.4 in\epsfclipon\epsfbox{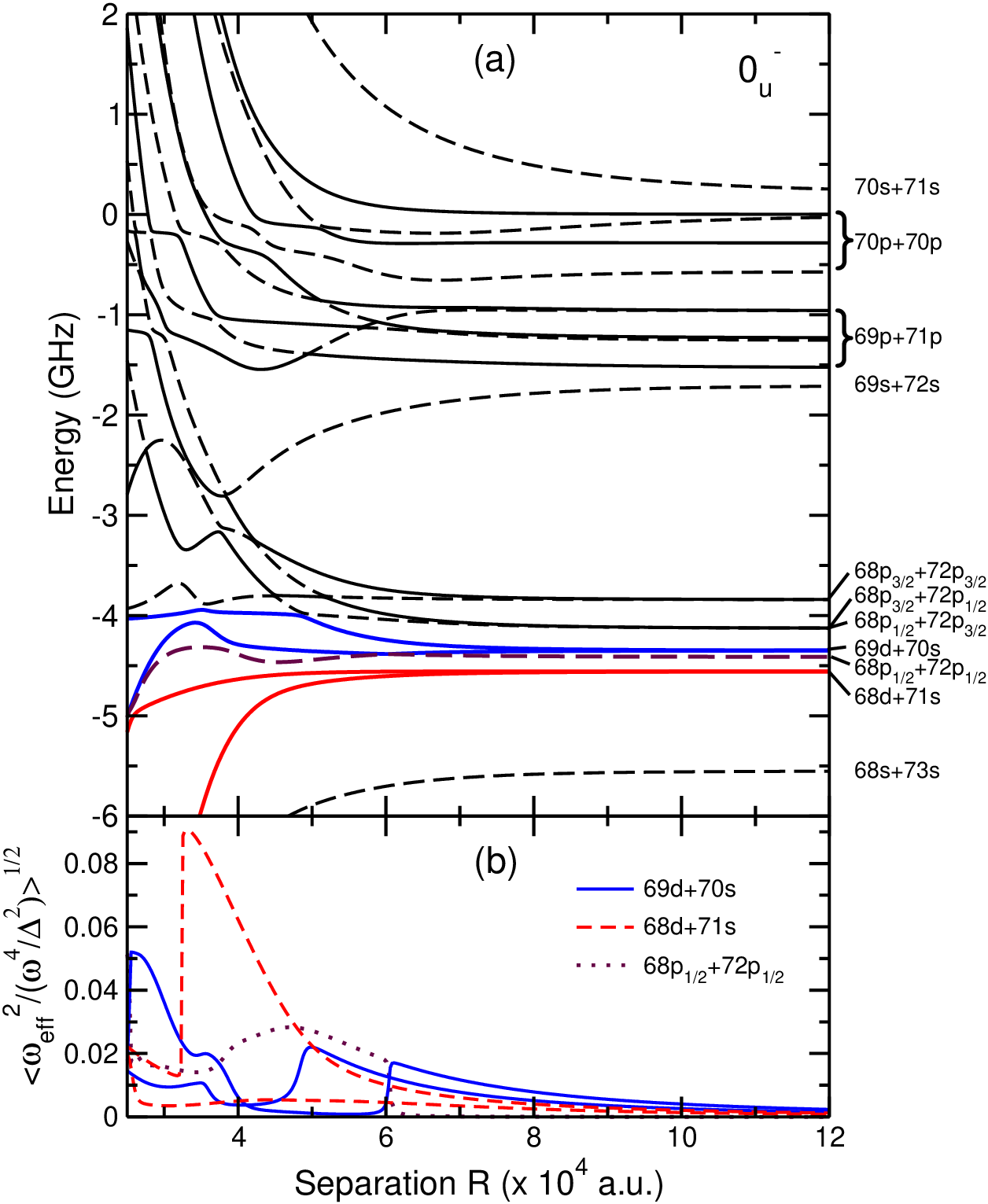}}
\caption{\label{0Usymm} Same as Fig. \ref{0Gsymm} but for the $0_u^{-}$ symmetry}
\end{figure}

We have assumed in this analysis that there is no background electric field. 
If a background electric field is included, then the group symmetry of the electronic Hamiltonian is not $D_{\infty h}$. Consequently, the quantum numbers based on this symmetry are not good anymore and we cannot reduce the interaction matrix in the basis of states of a given symmetry. In general, one has to include all possible basis states and use them to diagonalize the interaction matrix \cite{shaffer06}.

For non-overlapping charge distributions, the Rydberg-Rydberg interaction may be expanded in a series of inverse powers of nuclear separations $R$. In this paper, we consider dipolar and quadrupolar interactions only.  These interactions can be described by the following form \cite{buehler,marinescu97}
\begin{equation}\label{Vdd}
\begin{split}
   V_{L}(R) = & - \frac{(-1)^{\ell} 4\pi\, r_1^{L} r_2^{L}}{\hat{L}\, R^{2 L+1}}\\
  & \times\sum_{m} 
      B^{L+m}_{2\ell}  Y_{L}^{m}(\hat{r}_1)
      Y_{L}^{-m}(\hat{r}_2) , 
\end{split}
\end{equation}
where $L\!=\!1$ ($L\!=\!2$) for dipolar (quadrupolar) interactions, $B^m_n \equiv m !/(n-m) !$ is the binomial coefficient, $\vec{r}_i$ is the position of the electron $i$ around its center and $\hat{L} \!=\!2 L+1$.

Defining $|a\rangle|b\rangle \equiv |n_a, \ell_a, j_a ,m_a\rangle 
 |n_b, \ell_b,j_b ,\Omega-m_a \rangle$, one can show that the matrix elements of $ V_{\ell}(R)$ are
\begin{align}
& \langle 1 , 2 | V_L| 3, 4\rangle =(-1)^{L-1-\Omega+\sum\limits_{i=1}^4j_i} \sqrt{\hat{\ell}_1 \hat{\ell}_2 \hat{\ell}_3 \hat{\ell}_4      	\hat{j}_1  \hat{j}_2 \hat{j}_3 \hat{j}_4}    \nonumber\\
&\hspace{1.mm}\times\frac{\mathcal{R}_{13}^{(L)} \mathcal{R}_{24}^{(L)}}{R^{2L+1}}
	\left(\! \begin{array}{ccc}
     \ell_1 & L & \ell_3\\
     0&0&0
     \end{array} \! \right)      
    \left( \!\begin{array}{ccc}
     \ell_2 & L & \ell_4\\
     0&0&0
     \end{array}  \!\right) 
	\left\{\! \begin{array}{ccc}
       j_1 & L & j_3\\
     \ell_3&\frac12&\ell_1
     \end{array} \! \right\}
	 \nonumber\\ 
 &\hspace{1.mm}\times\left\{\! \begin{array}{ccc}
      j_2 & L & j_4\\
    \ell_4&\frac12&\ell_2
   \end{array} \! \right\}  \sum_{m=-L}^L  B^{L+m}_{2 L}\left(\! \begin{array}{ccc}
        j_1 & L & j_3\\
         -m_1&m&m_3
         \end{array} \! \right)\nonumber\\
 &\hspace{26.mm} \times
     \left(\!\begin{array}{ccc}
        j_2 & L & j_4\\
         -\Omega+m_1&-m&\Omega-m_3
         \end{array} \! \right) ,\label{matrix-element} 
\end{align}
where $\hat{\ell}_i=2\ell_i+1$, $\hat{j}_i=2j_i+1$, and $\mathcal{R}_{ij}^L$ is the radial 
part of the matrix element $\langle i|r^{L}|j\rangle$.

The asymptotic basis used to represent $U(R)$ consists of all $np\!+\!np$, 
$(n\!-\!1)d\!+\!ns$ and $(n\!-\!2)d\!+\!(n\!+\!1)s$ asymptotic states, as well as all the states in between with a significant coupling to these asymptotes. These include the $(n\!-\!1)p\!+\!(n\!+\!1)p$, $(n-1)s\!+\!(n+2)s$ and $(n\!-\!2)p\!+\!(n\!+\!2)p$ asymptotes. These states are our primary interest, but to describe them correctly at short internuclear distances one should also include other nearby states strongly coupled to them. Included nearby asymptotes  are $ns\!+\!(n+1)s$,  $(n-2)s\!+\!(n+3)s$, $(n\!-\!3)d\!+\!(n\!+\!2)s$, $n d\!+\!(n\!-\!1)s$, $(n\!-\!3)f\!+\!np$ and $(n\!-\!2)f\!+\!(n\!-\!1)p$. While the dipole-dipole interaction couples states belonging to different $n\ell+n'\ell'$ asymptotes, the quadrupole interaction is mainly relevant for states within the same $n\ell+n'\ell'$ asymptote. The only exceptions are the off-diagonal quadrupole matrix elements between $(n\!-\!1)d\!+\!ns$ and $(n\!-\!2)d\!+\!(n\!+\!1)s$ asymptotes. As mentioned before, the coupling depends on $|n_1-n_2|$, or to be precise, $|n_1^*-n_2^*|$, where the quantum defect $\delta_{\ell}$ is included in the effective principal quantum number $n_i^*$ as follows: $n_i^*=n_i-\delta_{\ell}$. Since the difference in effective principal quantum numbers of states $(n\!-\!1)d$ and $(n\!+\!1)s$, as well as $(n\!-\!2)d$ and $ns$ states, is only 0.22 for Rb (for high Rydberg states $\delta_{\ell=0}\approx 3.13$ and $\delta_{\ell=2}\approx 1.35$ \cite{gallagher03}), the off-diagonal quadrupole matrix element is several times larger than the diagonal ones. These asymptotes are very close in energy (separated by only 200 MHz for $n\!=\!70$) so that at $R\sim 30000$ $a_0$ and $n\!=\!70$, quadrupole off-diagonal coupling is comparable with the asymptotic energy spacing and these states become well mixed.
This off-diagonal quadrupole coupling is relevant for the shape of $(n\!-\!1)d\!+\!ns$ and $(n\!-\!2)d\!+\!(n\!+\!1)s$ resonances.
\begin{figure}
    \centerline{\epsfxsize=3.4 in\epsfclipon\epsfbox{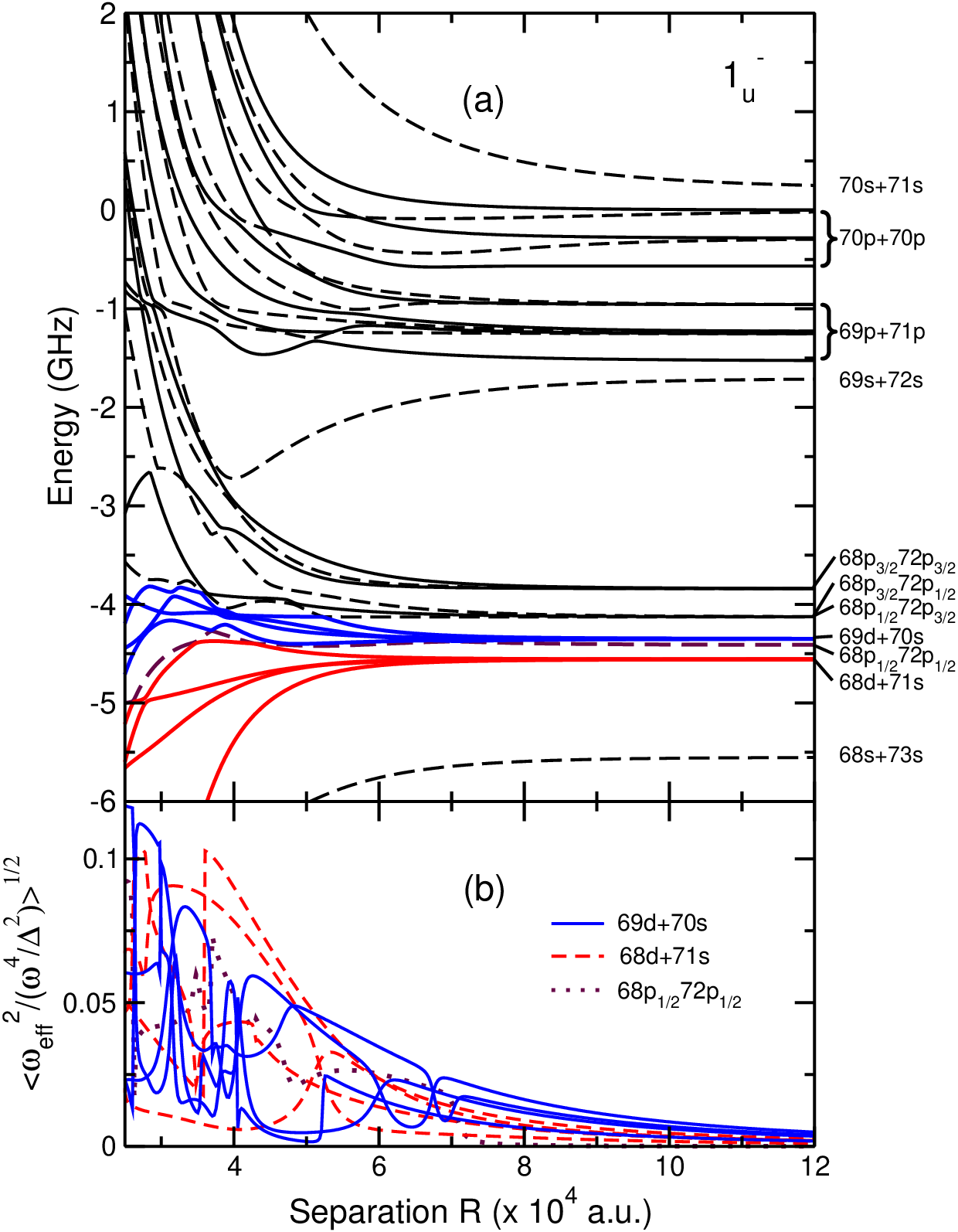}}
\caption{\label{1Usymm} Same as Fig. \ref{0Gsymm} but for the $1_u$ symmetry, except that each  $nd+n's$ asymptote has four states of this symmetry.}
\end{figure}

\subsection{Excitation of a pair of interacting Rydberg atoms}
We treat the $(n-1)d+ns$ molecular resonances as two-body phenomena, which is supported by the fact that their energies coincide with the average value of only two atomic energies \cite{farooqi03}. The resonances  are far red-detuned, so normally there should not be many excited atoms and presumably pair-wise excitation is the dominant mechanism. The general problem to be solved is the same as in \cite{Stanojevic} for $(n-1)p+(n+1)p$ resonances. However, the analysis of $(n-1)d+ns$ resonances is technically much more demanding and some approximations we have applied before may not be satisfactory here. To check this, we have used several different ways to evaluate the contributions to the number of excited atoms from various  molecular states. We discuss them in the results section.

We consider a two-body Hamiltonian that includes long-range interactions and a linearly polarized optical field. For simplicity, we include only one molecular state  $|\varphi_{\lambda}\rangle$ and the corresponding molecular potential $\epsilon_{\lambda}(R)$. The Hamiltonian is (in the rotating frame and in the rotating-wave approximation)
\begin{align}\label{hamiltonian}
 H =  \sum_{i=1}^2 &\left[\Delta\sigma_{ee}^i+\Delta'\sigma_{e'e'}^i \right]+\sum_{i=1}^2 \left[\frac{\omega}{2}\sigma_{eg}^i   +\frac{\omega'}{2}\sigma_{e'g}^i + \mathrm{h.c.}\rule{0mm}{4 mm}\right]\nonumber\\
    & +\left[\Delta_{\lambda}+\epsilon_{\lambda}(R)\right]
         |\varphi_{\lambda}\rangle\langle\varphi_{\lambda}|\; ,
\end{align}
where $\omega$, $\omega'$, and $\Delta$, $\Delta'$ are single-photon Rabi 
frequencies and detunings relative to the $np_{3/2}$ and $np_{1/2}$ fine-structure components, respectively. Here $\Delta_{\lambda}$ is the two-photon detuning from the 
asymptotic $\epsilon_{\lambda}(R\rightarrow\infty)$ molecular level. The operators $\sigma_{eg}^i$ 
and $\sigma_{ee}^i$ are defined as follows: $\sigma_{eg}^i \!=\!  \sum_m |e_i,m\rangle\langle g_i,m|$ and $\sigma_{ee}^i \!=\! \sum_m |e_i,m\rangle\langle e_i,m|$. To calculate $\omega$ and $\omega'$ we use the experimental values for the oscillator strength $f_{3/2}$ \cite{Shabanova} and the ratio \cite{Tong} $\omega/\omega'=\sqrt{f_{3/2}/f_{1/2}}\approx2.3$, which reflects the non-statistical character of the oscillator strengths $f_{3/2}$ and $f_{1/2}$.  

The excitation process is essentially a three-level scheme, although the number of states involved in the process is greater than three. We assume that a pair of ultracold Rb atoms is initially in one of the $5s+5s$ ground states. There are four possible  ground states $|5s,m_j\rangle|5s,m'_j\rangle$, corresponding to different projections of spin $m_j=\pm 1/2$. Ultimately, the total probability is averaged over all possible initial states. In the first excitation step, one of the atoms is excited to a given $np_j$ state. There are two intermediate states if the ground-state atoms have the same projections $m_j$, otherwise there are four of them. These intermediate states are further excited in the second step. The final state in this excitation scheme can be a single molecular state $|\varphi_{\lambda}(R)\rangle$ or a superposition of states. We consider both cases but in the equations a single  molecular state is assumed. Including more states in the equations is very straightforward and there is no need to do it explicitly. Eventually, to get the final excitation probability per atom for a given optical frequency, the contributions from all molecular potentials and all atom pairs that include a given atom are collected.

To get pair excitation probabilities we solve the time-dependent Schr\"odinger equation for the ground diatomic state, all intermediate states and a given doubly-excited molecular state $|\varphi_{\lambda}\rangle$. Here we present in detail the treatment  if two $m_j$ of the diatomic ground state are different and give only the final result when they are the same (this case is considered in detail in \cite{Stanojevic}).  Utilizing the fact that symmetric and antisymmetric states have independent time evolutions, the first step is to construct symmetric and antisymmetric combinations
\begin{equation}\label{symm-states}
\!|ij\rangle\!=\! \frac{1}{\sqrt{2}}
 \left\{\left|i,m\right>\left|j,-m\right> \!+\!q \left|i,-m\right>\left|j, m\right>\right\},
\end{equation}
where $q=1(-1)$ for symmetric(antisymmetric) states and  $m=1/2$ for linear laser polarization. In this way the diatomic ground state
$|gg\rangle$, four intermediate states $|ge\rangle$, $|eg\rangle$, $|ge'\rangle$, $|e'g\rangle$, and four doubly-excited states $|ee\rangle$, $|ee'\rangle$, $|e'e\rangle$, $|e'e'\rangle$ are defined. Here $e$ and $e'$ refer to $np_{3/2}$  and $np_{1/2}$ states, respectively. If there were no interactions, diatomic states $|ee\rangle$, $|ee'\rangle$, $|e'e\rangle$, $|e'e'\rangle$ would be the states directly accessible in two-photon excitation.  Any molecular state
$|\varphi_{\lambda}\rangle$ is accessible if it has some components of these diatomic states. Due to $\ell$-mixing induced by interactions, many $|\varphi_{\lambda}(R)\rangle$ gain a significant $np$ fraction at some finite internuclear separation $R$.

The wave function is modeled as follows
\begin{equation}\label{wave_f}
\begin{split}
\left|\psi\right>=&c_0 \left|gg \right>+c_{11}\left|ge \right> +c_{12}\left|eg \right> +c'_{11}\left|ge' \right> \\
&{}+c'_{12}\left|e'g \right> +c_2 \left|\varphi_{\lambda}\right>.
\end{split}
\end{equation}
Solving the time-dependent Schr\"odinger equation $i \partial \psi/\partial t=H \psi$ ($\hbar=1$) leads to the coupled system for the excitation amplitudes $c(t)$, 
\begin{align}
&\hspace{-2mm}i\frac{dc_0}{dt} \!= \! \frac{\omega^{*}}{2}(c_{11}+c_{12})
                  +\frac{\omega'^{*}}{2}(c'_{11}+c'_{12}), \label{Eqs-system-1}\\	
&\hspace{-2mm}i\frac{dc_{11}}{dt}\!=\! \Delta c_{11}\!+\! \frac{\omega}{2}c_0\!+\!\frac{\omega^{*}}{2}
	\langle ee |\varphi_{\lambda}\rangle c_2 \!+\!\frac{{\omega'}^{*}}{2}\langle e'e |\varphi_{\lambda}\rangle c_2,\label{Eqs-system-2}\\
&\hspace{-2mm}i\frac{dc_{12}}{dt} \!= \! \Delta c_{12} \!+\! \frac{\omega}{2}c_0
                 \!+\!\frac{\omega^{*}}{2}\langle ee |\varphi_{\lambda}\rangle c_2 \!+\!\frac{\omega'^{*}}{2}\langle ee' |\varphi_{\lambda}\rangle c_2, \label{Eqs-system-2}\\
      &\hspace{-3mm}i\frac{dc'_{11}}{dt}\!= \! \Delta c'_{11} \!+\! \frac{\omega'}{2}c_0 \!+\!\frac{\omega^{*}}{2}\langle ee'  |\varphi_{\lambda}\rangle c_2\!+\!\frac{\omega'^{*}}{2}\langle e'e' |\varphi_{\lambda}\rangle c_2,
 \label{Eqs-system-4}\\
&\hspace{-2mm}i\frac{dc'_{12}}{dt} \!= \! \Delta c_{12} \!+\! \frac{\omega}{2}c_0
                   \!+\!\frac{\omega^{*}}{2}\langle ee |\varphi_{\lambda}\rangle c_2 \!+\!\frac{\omega'^{*}}{2}\langle ee' |\varphi_{\lambda}\rangle c_2,\label{Eqs-system-5}\\
&\hspace{-2mm}i\frac{dc_2}{dt}\!=\! \left(\Delta_{\lambda}\!+\! \epsilon_{\lambda}(R)\right)c_2 +\Big(\frac{\omega}{2}\langle \varphi_{\lambda}|ee \rangle  +\frac{\omega'}{2}\langle\varphi_{\lambda}|e'e\rangle\Big)c_{11}\nonumber\\
              &\hspace{2mm} +\Big(\frac{\omega}{2}\langle\varphi_{\lambda}|ee\rangle \!+\!\frac{\omega'}{2}\langle\varphi_{\lambda}|ee'\rangle\Big)c_{12}\label{Eqs-system-6}+\Big(\frac{\omega}{2}\langle\varphi_{\lambda}|ee'\rangle \\
	&\hspace{2mm}{}+\!\frac{\omega'}{2}\langle \varphi_{\lambda}|e'e'\rangle\Big)c'_{11}
                +\left(\frac{\omega}{2}\langle\varphi_{\lambda}|ee\rangle +\frac{\omega'}{2}\langle \varphi_{\lambda}|ee'\rangle\right)c'_{12} . \nonumber
\end{align}
The analogous system of equations for $m_j=m'_j=1/2$ was considered in detail \cite{Stanojevic} in the analysis of $(n-1)p+(n+1)p$ resonances. The projections onto the molecular state are defined as: 
$a_{ee}(\lambda)=\langle ee|\varphi_{\lambda}\rangle$, 
$a_{ee'}(\lambda)=\langle ee'|\varphi_{\lambda}\rangle$, 
$a_{e'e}(\lambda)=\langle e'e|\varphi_{\lambda}\rangle$, 
and $a_{e'e'}(\lambda)=\langle e'e'|\varphi_{\lambda}\rangle$.
We assume that all of the $a_{ij}$ coefficients are real.

One can obtain a tractable system after the elimination of the excitation amplitudes of all intermediate states. This is justified by the fact that the dominant frequencies governing
their time evolutions are $\Delta$ and $\Delta'$ which are much larger than the relevant Rabi frequencies. On a molecular resonance, $\Delta$ and $\Delta'$ are about $2\pi \cdot 2.2$ GHz, while the peak values of $\omega$ and $\omega'$ are about 2.3 GHz and 1 GHz, respectively (for the actual experimental parameters). Laser intensities used in this experiment were almost two order of magnitude higher than the ones used for the $(n-1)p+(n+1)p$ resonances so that we have to check for the effects of higher laser fields. We adiabatically eliminate $c_{11}$, $c_{12}$, $c'_{11}$ and $c'_{12}$, as in \cite{Stanojevic}, but this time keeping power-dependent terms. The result is equivalent to the Bloch equations of a two-level system
\begin{align}
&i\frac{dc_0}{dt} =-\omega_{g}(t) c_0 -\frac{\omega_{\mathrm{eff}}^*}{2}c_2, \label{blck-eqs-1}\\
&i\frac{dc_2}{dt} = \left[\Delta_{\lambda}+ \epsilon_{\lambda}(R)-\omega_{e}(t)\right]c_2 
                 -\frac{\omega_{\mathrm{\mathrm{eff}}}}{2}c_0,\label{blck-eqs-2}
\end{align}
where the effective two-photon Rabi frequency is 
\begin{equation}\label{ef-Rabi}
\begin{split}
\omega_{\rm eff}=&\frac{\omega^2}{\Delta} a_{ee}(\lambda)
                 +\frac{\omega\, \omega' }{2} 
           \frac{\Delta + \Delta'}{\Delta \, \Delta'} (a_{ee'}(\lambda)+a_{e'e}(\lambda)),\\
& {}+\frac{\omega'^2}{\Delta'} a_{e'e'} (\lambda)
\end{split}
\end{equation} 
and the power-dependent terms $\omega_{g}$ and $\omega_{e}$ are 
\begin{align}
&\omega_{g}= \frac{|\omega|^2}{2\Delta}+\frac{|\omega' |^2}{2\Delta'},\label{ACg}\\
 &  \omega_{e} = \frac{|\omega a_{ee}+\omega'a_{e'e}|^2}{4 \Delta}+\frac{|\omega a_{ee}+\omega'a_{ee'}|^2}
{4 \Delta}\nonumber\\
 &\hspace{8mm} + \frac{|\omega'a_{e'e'}+\omega a_{ee'}|^2}{4\Delta'} + \frac{|\omega'a_{e'e'}+\omega a_{e'e}|^2}{4\Delta'}.\label{ACe}
\end{align}
These $\omega_{g}(t)$ and $\omega_{e}(t)$ can be interpreted as power dependent shifts of the diatomic ground and doubly-excited states, respectively. Also, $\omega_{g}$ is much greater than $\omega_{e}$ because it does not depend on $a_{ij}$, which measure the $p$-character of the doubly-excite state. For the experimental parameters, the peak value of $\omega_{g}(t)$ on a molecular resonance is about 280 MHz, which is comparable with the laser bandwidth. 

In the vicinity of the molecular resonance, the second term on the right-hand side of Eq. (\ref{blck-eqs-1}) is much smaller than the first one, and thus can be ignored. After neglecting that term, Eq. (\ref{blck-eqs-1}) can be solved
\begin{equation}\label{c_0Solut}
 c_0(t) =\exp\left[i\int_{t_0}^t\omega_{g}(t')\,dt'\right].   
\end{equation}
This $c_0$ can be used to find $c_2$ after the phase transformation $c_2\equiv\exp[-i(\Delta_{\lambda}+\epsilon_{\lambda}(R))t -\int_{t_0}^t\omega_{e}(t')\,dt' ]\bar{c}_2$ is performed.
The excitation amplitude $\bar{c}_2$ is
\begin{equation}\label{c_2Solut}
\begin{split}
\bar{c}_2 =i \int_{t_0}^t& dt_1 \frac{\omega_{\mathrm{eff}}(t_1)}{2}\\ &\times\;e^{i(\Delta_{\lambda}+\epsilon_{\lambda}(R))t_1+\int_{t_0}^{t_1}(\omega_{g}(t_2)-\omega_{e}(t_2))dt_2 }. 
\end{split}
\end{equation}
We assume that excitation laser pulses have a Gaussian time profile. It has been shown that the excess bandwidth in this experimental setup is mainly due to a linear frequency chirp. Because of the constant ratio $\omega/\omega'$, both $\omega$ and $\omega'$ have the same time dependence $\omega,\omega'\sim\exp(-(1+i \gamma) t^2/\sigma^2$, where $\sigma$ measures the pulse duration and $\gamma$ is a chirp parameter related to the laser bandwith $\Gamma$ and duration $\sigma$ as follows
\begin{equation}\label{beta}
 \gamma =\sqrt{ \frac{\pi^2 \Gamma^2 \sigma^2-2\ln2}{2\ln2}  }.
\end{equation}

The probability to excite $|\varphi_{\lambda}(R)\rangle$ from the initial state $|5s,m_{1/2}\rangle|5s,m_{1/2}\rangle$ is $ P_1(\lambda) = |\bar{c}_2(t\rightarrow\infty)|^2$. 
According to Eqs. (\ref{ACg}-\ref{ACe}), $\omega_{g}$ and $\omega_{e}$ do not depend on the chirp. If $\omega_{g}$ and $\omega_{e}$ in Eq. (\ref{c_2Solut}) can be ignored, a simple formula can be derived \cite{Stanojevic}. 

\begin{equation}\label{probab11}
|\bar{c}_2|^2 = \frac{\beta^2(\lambda) \pi}{2} \frac{I^2}{I^2_{\rm sat}}\frac{\pi\ln^2 2}{\tau^3 \Gamma} 
               2^{-[\Delta_{\lambda}+ \epsilon_{\lambda}(R)]^2/2\pi^2 \Gamma^2},
\end{equation}
where $I_{\mathrm{sat}}$ is the saturation intensity for ideal unchirped light and isolated $np_{3/2}$ atoms and $\tau$ is a Gaussian pulse duration (FWHM). The approximation applied to get this formula is equivalent to neglecting the $\omega_e$-term in Eq. (\ref{blck-eqs-2}) and both terms in the right-hand side of Eq. (\ref{blck-eqs-1}) (i.e. using $c_0=1$). In Eq. (\ref{probab11}), $\beta(\lambda)$ is a time-independent part of $\omega_{\rm eff}$ defined via $\omega_{\rm eff}(t)=\beta(\lambda)\omega^2(t)$. 

If all $a$-coefficients are equal to zero, according to Eq. (\ref{ef-Rabi}), the effective two-photon Rabi frequency is also zero. These coefficients measure different $p$-characters in 
$|\varphi_{\lambda}(R)\rangle$. To evaluate them, we have to express the doubly-excited states $|ee'\rangle$, $|ee\rangle$, and $|e'e'\rangle$ (defined in the 
space-fixed frame) in the molecule-fixed reference frame, where all $|\varphi_{\lambda}(R)\rangle$ are naturally defined \cite{Stanojevic}. Apparently, all angular dependence related to different orientations of the molecular axis is contained in these $a$-coefficients, and the observable quantities have to be averaged over all spatial orientations of the internuclear axis.   
\begin{figure}[b]
    \centerline{\epsfxsize=3.3in\epsfclipon\epsfbox{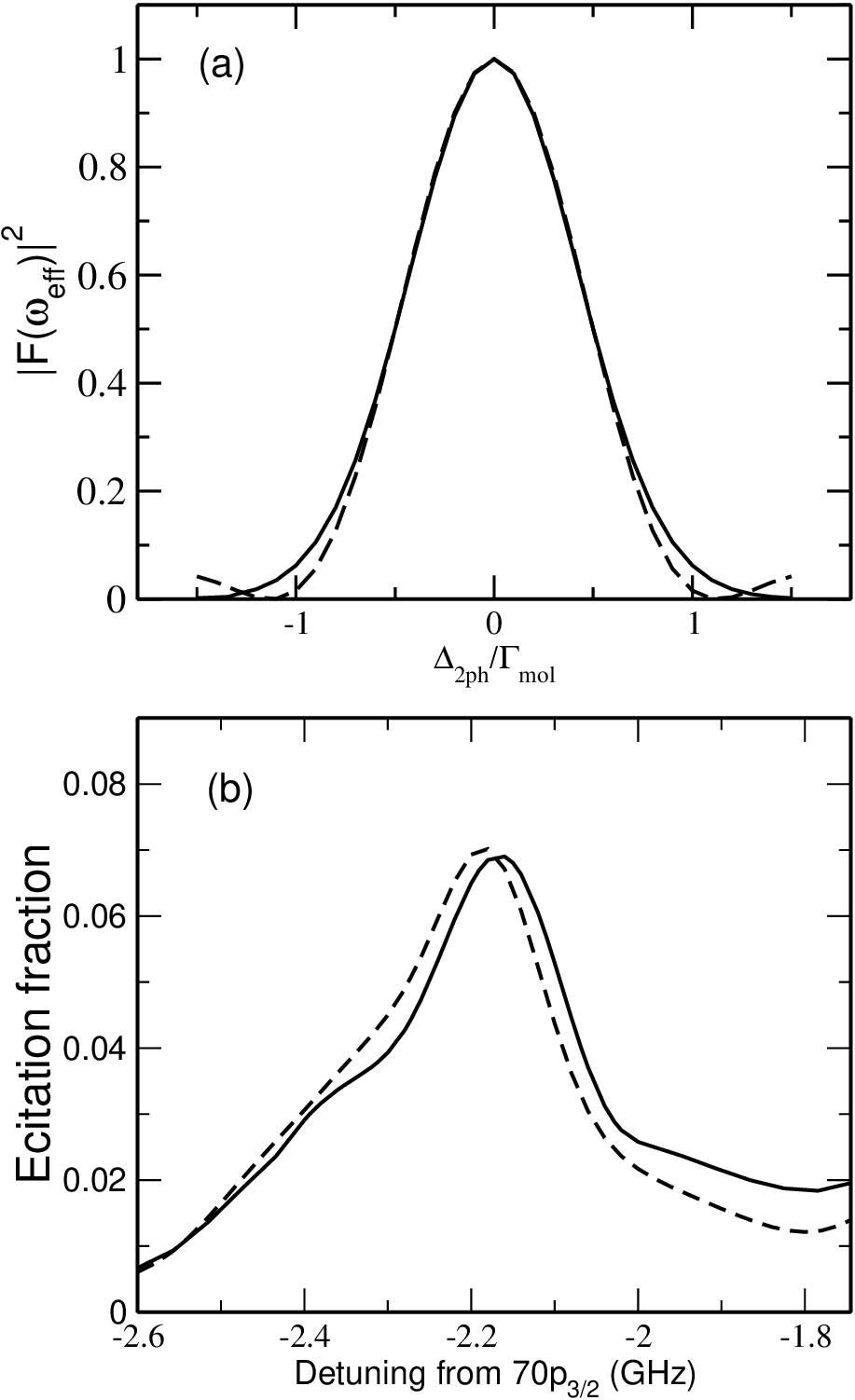}}
\caption{\label{methods} (a) Fourier transforms of $\omega_{\rm 2ph}(t)$ of a chirped Gaussian pulse (dashed line) and its counterpart square pulse (solid line). The pulse areas and $\Gamma_{\rm 2ph}$ (FWHM) of $\omega_{\rm 2ph}(t)$ for both pulses are matched to minimize the difference in the excitation probabilities. The tail of the square pulse vanishes much slower so it is truncated at $\Delta_{\rm 2ph}/\Gamma_{\rm 2ph}=1.6$, where $\Delta_{\rm 2ph}$ is the two-photon detuning. The remaining small difference between the pulses  is expected to diminish further after summing over all atom pairs in (\ref{eq:Pexc_estim}). (b) Comparison between  calculated molecular signals using Eq. (\ref{probab11}) (dashed line) and Eq. (\ref{truncate_solution}) (solid line). The latter is based on the exact solution for a square pulse assuming that, in general, superpositions of molecular states are excited. The differences are somewhat larger just above the molecular resonance, where there are many very close molecular potentials, as shown in Figs. \ref{0Gsymm}-\ref{1Usymm}. The assumed laser bandwidth in this plot is 200 MHz.}
\end{figure}

For the case $m_j=m'_j=\pm1/2$, the ground and doubly-excited states are respectively  $|gg\rangle =|g,m\rangle|g,m\rangle$ and  $|ee\rangle=|e,m\rangle|e,m\rangle$. Intermediate states $|ge\rangle$, $|ge'\rangle$
and doubly-excited state $|ee'\rangle$ have the form $|ij\rangle= \frac{1}{\sqrt{2}}
\left\{|i,m\rangle|j,m\rangle + |j,m\rangle|i, m\rangle\right\}$. The coefficients $a_{ij}$ are defined as before. After the elimination of excitation amplitudes of all intermediate states, one gets again the system (\ref{blck-eqs-1}-\ref{blck-eqs-2}) with different $\omega_{\rm eff}(t)$ and $\omega_{e}(t)$
\begin{equation}\label{eff-Rabi2}
 \omega_{\rm eff}=\frac{\omega^2}{\Delta} a_{ee}(\lambda)
              \! +\!\frac{\omega'^2}{\Delta'} a_{e'e'}(\lambda)
	{}+\frac{\omega\, \omega' }{\sqrt{2}}
        \frac{\Delta \!+\! \Delta'}{\Delta \, \Delta'} a_{ee'}(\lambda),
\end{equation} 
\begin{equation}\label{ACe2}
\omega_{e} = \frac{|\sqrt{2}\omega a_{ee}+\omega' a_{ee'}|^2}{4\Delta}
   +\frac{|\sqrt{2}\omega' a_{e'e'}+\omega a_{ee'}|^2}{4\Delta'}.
\end{equation}
The solutions for these initial states are also given by Eq. (\ref{c_2Solut}), or for low laser intensity by 
Eq. (\ref{probab11}). These excitation probabilities must also be averaged over all spatial orientations of the internuclear axis. In (\ref{probab11}) this is applied to  $\beta(\lambda)$  only.

For an unpolarized sample of ultracold atoms, all initial states are equally probable, so that excitation probabilities have to be averaged over all initial diatomic states and then all contributions to a given $|\varphi_{\lambda}(R)\rangle$ are summed up.  This is done for many $|\varphi_{\lambda}\rangle$ to include the possibility of exciting different molecular potentials. The result is the averaged excitation $P_2 (R)$ of an atom pair. The excitation probability per atom is the sum of all excitation 
probabilities of atom  pairs that include a given atom,
\begin{equation}\label{probab-av}
 P_{\mathrm{exc}}  = 4\pi 
    \int_{0}^{\infty}dR\,R^2 \rho \; P_2 (R),
\end{equation} 
where $\rho$ is the sample density. 

\begin{figure}[t]
    \centerline{\epsfxsize=3.4 in\epsfclipon\epsfbox{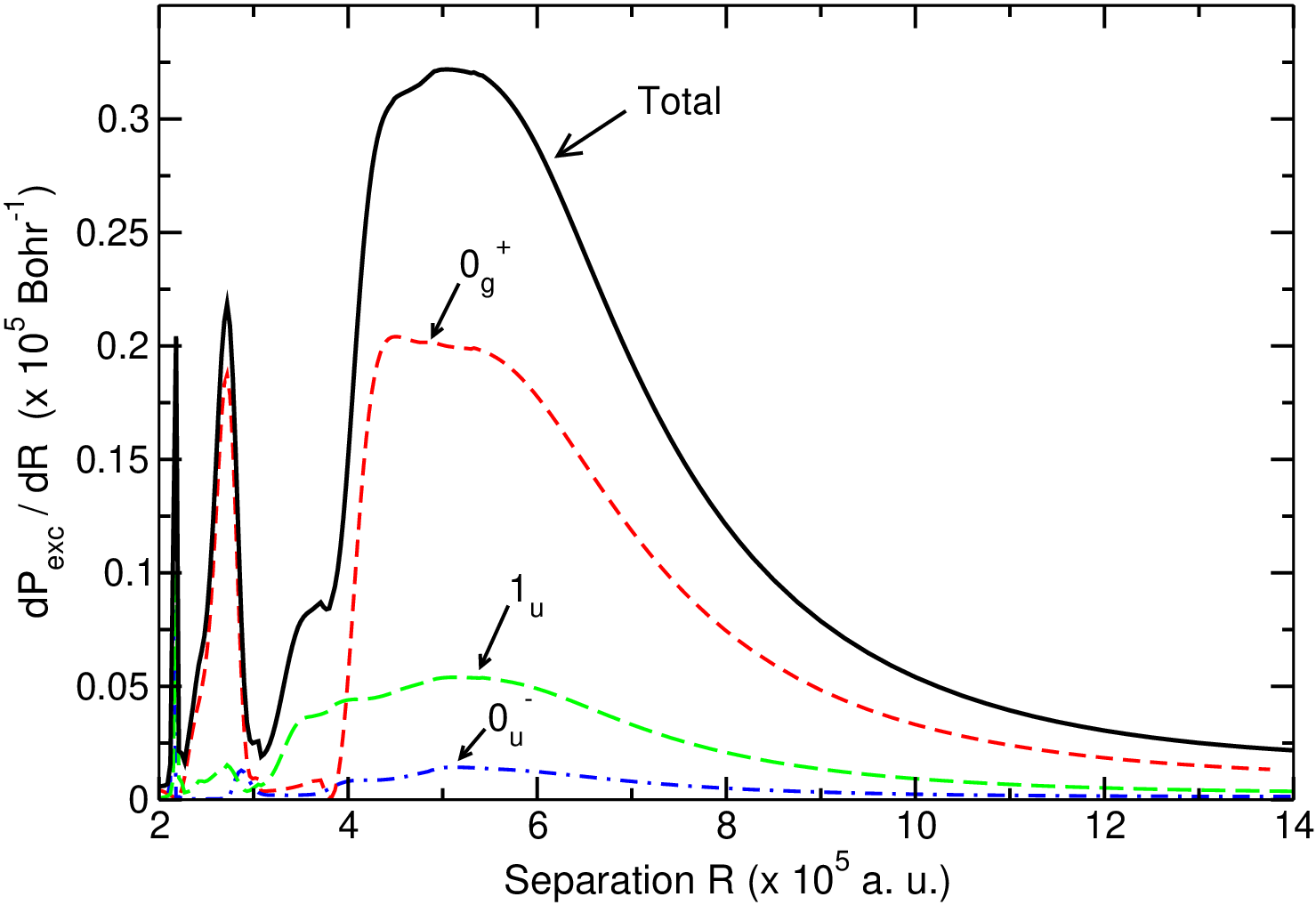}}
\caption{\label{radialProb} Pair excitation probability as a function of separation $R$. We show contributions from all states associated with the three most relevant asymptotes. The total dependence includes twice the contribution of $1_u$ since this state is two-fold degenerate. According to Eq. (\ref{probab-av}), $dP_{\rm exc}/dR=4\pi \rho R^2 P_2 (R)$,
where $P_2 (R)$ is the excitation probability of an atom pair. Note that the weighted factor $R^2$ is included in the shown dependence. This figure shows that pairs at shorter distances are difficult to excite, even though $\ell$- mixing and their $\langle \omega_{\rm eff}^2\rangle$ is much greater there (Figs. \ref{0Gsymm}-\ref{1Usymm}) because they are very detuned from the resonance. The main contribution comes from pairs at separations $R\geq 40000$ $a_0$.}
\end{figure}

Molecular potentials can be very close to each other for some $R$, as shown in Figs. 1-3. For such $R$ atom pairs are excited into  superpositions of molecular states.  Altough the extension of the system (\ref{Eqs-system-1}-\ref{Eqs-system-6}) to include more molecular states $|\varphi_{\lambda}\rangle$ is simple, the technical difficulties of solving it are not negligible. Not only do we have four parameters to vary ($R$, $I$, laser frequency and the orientation of the molecular axis), but we also have many molecular potentials and several symmetry cases. Also, for a given optical frequency, different molecular states are excited at different distances $R$. This is especially true at short $R$ for which the potentials vary significantly. 
Instead of doing this, we choose a different approach to account for possible superpositions of molecular states. This approach is in many ways as complete as the full numerical calculation that includes the full set of doubly-excited states and at the same time is no more difficult than the calculation of molecular potentials itself. As Eq. (\ref{probab11}) suggests, only a few parameters related to the excitation laser, such as the bandwidth and pulse duration, are important. The details of the laser pulse cannot be of fundamental importance, so we can substitute for the actual chirped Gaussian pulse a square pulse with the parameters chosen to give probabilities consistent with (\ref{probab11}). We have to match the pulse area and the width (actually FWHM) of the Fourier spectrum corresponding to the two-photon Rabi frequency $\omega_{\rm eff}$. Therefore, $\Gamma$ and $\tau$ are not our direct concern, but rather $\Gamma_{\rm 2ph}$ and $T_{\rm 2ph}$, which characterize $\omega_{\rm eff}(t)$.  For a Gaussian pulse $\Gamma_{\rm 2ph}=\sqrt{2}\Gamma$. This $\Gamma_{\rm 2ph}$ would also be the FWHM of a square pulse if  its duration were chosen to be $T_{\rm sq}=2.783/\pi \Gamma_{\rm 2ph}$.  The single-photon Rabi frequency $\omega_{\rm sq}$ of the square pulse is chosen to provide equal pulse areas of $\omega_{\rm eff}(t)$ of the actual pulse and its substitute. A great adventage of this approach is that now we can easily write and calculate the exact excitation probabilities of all molecular states. The total Hamiltonian $H_{\rm tot}$ of the system consists of the long-range interaction part $U(R)$, given by Eq. (\ref{Utotal}) and the optical field ($\hbar=1$).
\begin{equation}\label{hamilton_tot}
\begin{split}
 H_{\rm tot}(R) =& U(R)+\sum_{i=1}^2 \left[\Delta\sigma_{ee}^i+\Delta'\sigma_{e'e'}^i \right]\\
        &{}\hspace{-0.1in}+\sum_{i=1}^2 \left[\frac{\omega_{\rm sq}}{2}\sigma_{eg}^i 
        +\frac{\omega'_{\rm sq}}{2}\sigma_{e'g}^i +  \mathrm{h.c.}\right].
\end{split}
\end{equation}
To represent $H_{\rm tot}$ we use the basis of $U(R)$ completed by the intermediate states and the ground diatomic states. The matrix elements of $H_{\rm tot}$, corresponding to these added basis states, are essentially given by Eqs. (\ref{Eqs-system-1}-\ref{Eqs-system-6}). The only modification is to replace $|\varphi_{\lambda}\rangle$ by a superposition of different $|\varphi_{\lambda}\rangle$. The matrix of $H_{\rm tot}$ is only slightly bigger than the matrix of $U(R)$, so solving the eigenproblem  of $H_{\rm tot}$ does not impose additional difficulties. If $\varepsilon_{i}(R)$ are the eigenvalues related to the eigenvectors $\left|\phi_{i}(R)\right>$ of $H_{\rm tot}(R)$, then the solution of the time-dependent Schr\"odinger equation $i \partial \left|\psi\right>/\partial t=H_{\rm tot} \left|\psi\right>$ is 
\begin{equation}\label{solution_tot}
\left|\psi(t;R)\right> =\sum_{i}\! \left<\phi_{i}(R)|gg\right>
	 e^{-i\varepsilon_{i}(R)t}\left|\phi_{i}(R)\right>.
\end{equation}
Unlike Eqs. (\ref{c_2Solut},\ref{probab11}), the last formula does not give unphysical probabilities for large laser power. In the actual calculation we use a modification of Eq. (\ref{solution_tot}) to account for the difference in excitation probabilities for large detunings.  Excitation probabilities  for a square pulse do not vanish sufficiently fast. This happens because the tails of the Fourier spectra of these two types of pulses are very different at large detunings. This is illustrated in Fig. \ref{methods}(a).  We overcome this by truncating the contributions from large detunings  
\begin{equation}\label{truncate_solution}
\begin{split}
\left|\psi_e(t;R)\right>  = \sum_{i}&\Theta(|\varepsilon_{i}(R)-2\Delta|-\eta \Gamma_{\rm mol}) \\
	&\times\left<\phi_{i}(R)|gg\right> e^{-i\varepsilon_{i}(R)t}\left|\phi_{i}(R)\right>, 
\end{split}
\end{equation}
where $\Theta$ is the Heaviside function and $\eta$ is the cut-off parameter (our choice $\eta=1.6$ is justified by  Fig. \ref{methods}(a)). We note that the radial dependence, shown in Fig. \ref{radialProb}, is obtained using the last equation. One can find the same dependence using the simpler formula (\ref{c_2Solut}). Both ways give essentially the same radial dependence.   In the last formula all $\varepsilon_{i}(R)$ are expressed with respect to the energy of asymptotic $np_{3/2}+np_{3/2}$ states.

\subsection{$n$-scaling}
Now we estimate the major contribution to the $n$-scaling of the $(n\!-\!1)d+ns$ resonance signal. For this purpose we use a rather simplified description of these states and the resonances. First we ignore fine structure. As Fig. \ref{radialProb} suggests, the major contribution to the molecular signal does not come from the region of strong $\ell$-mixing, but rather from the long-range region with considerably weaker mixing. In this estimate we completely neglect the contribution of strong $\ell$-mixing at short $R$ to the molecular signal. The wave function of $(n\!-\!1)d+ns$ states can be expanded as follows 
\begin{equation}\label{eq:ds_state}
\begin{split}
 |(n\!-\!1)dns;R\rangle=&|(n\!-\!1)dns^{(0)}\rangle+\alpha_{pp}(R)|npnp^{(0)}\rangle\\
& {}+ |\varphi_{res};R\rangle \; ,
\end{split}
\end{equation}
where $|\varphi_{res};R\rangle$ is the residue of the expansion and $|ndn's^{(0)}\rangle$, $|npnp^{(0)}\rangle$ are asymptotic states $|(n\!-\!1)dns;R\!\rightarrow\!\infty\rangle$ and $|npnp;R\!\rightarrow\!\infty\rangle$. In general, there are more than one $|np\,np^{(0)}\rangle$ state in the last expansion but the scaling law, in this approximation, does not depend of their number so keeping only one term is sufficient for our purpose. We want to find the function $\alpha_{pp}(R)$ because the two-photon Rabi frequency is directly proportional to it. The Hamiltonian is still given by Eq. (\ref{Utotal}). In the first approximation $\alpha_{pp}(R)$ is
\begin{equation}\label{eq:alpha}
\alpha_{pp}(R)=\frac{  \langle np\,np^{(0)}|V_{Ryd}(R)|nd\,n's^{(0)}\rangle   }{ Eds_{0}-Epp_{0} },
\end{equation}
where $Eds_{0}-Epp_{0}$ is the asymptotic energy spacing of diatomic $(n\!-\!1)d+ns$ and $np+np$ levels. The last formula is valid in the region of weak $\ell$-mixing for $(n\!-\!1)d+ns$ states. We conclude that
\begin{equation}\label{eq:alpha_scaling}
\alpha_{pp}(R)\sim n^7/R^3,
\end{equation}
since only the dipole-dipole part of $V_{Ryd}(R)$ couples those asymptotic states, and $Eds_{0}-Epp_{0}\sim n^{-3}$.

We proceed using the results of the previous sections, but ignoring many details which are not of great importance 
for $n$-scaling. The $n$-scaling is well defined only if there is no saturation of excitation so that the two-photon absorption probability $P_{2}$ per pair is
\begin{equation} \label{eq:P-2}
   P_{2}\sim |\omega_{{\rm eff}}|^2  \; .
  \end{equation}
We have $\omega_{\rm eff}\sim\omega^2\alpha_{pp}(R)/\Delta$, where $\omega$ is the single-atom Rabi frequency defined in the previous sections and $\Delta\approx (Eds_{0}-Epp_{0})/2$ is the detuning from the atomic $np$ resonance. Note that $\omega^2/\Delta$ is in the first approximation $n$-independent because the single photon
Rabi frequency $\omega$ scales as the dipole matrix element $ n^{-3/2}$ and $\Delta\sim n^{-3}$.

To get the excitation probability $P_{{\rm exc}}$ we use Eq. (\ref{eq:P-2})
\begin{equation} \label{eq:Pexc_estim}
P_{{\rm exc}}\sim\frac{\omega^4 }{\Delta^2} \int_{R_0}^{\infty}dR\,R^2 |\alpha_{pp}(R)|^2
\sim n^{14}/R_0^3.
  \end{equation}
The lower limit $R_0$ is set as follows. We assume that the laser frequency corresponds to the two-photon resonance, whose position coincides, to a very good approximation, with the asymptotic energy of the $(n-1)d+ns$ state. A pair of atoms will be out of two-photon resonance if its interaction energy $U_{ds}(R_0)$ is greater than $\Gamma_{2ph}\sim \Gamma$, therefore $|U_{ds}(R_0)|\sim \Gamma$. Because the laser bandwidth is considerably narrow (close to the Fourier transform limit), 
$U_{ds}(R_0)$ is not very large, and assuming weak $\ell$-mixing, we can use second order perturbation theory to find an estimate for it. Eventhough the $(n-2)p+(n+2)p$ states are very close to  $(n-1)d+ns$ states, they are very weakly coupled to them so they basically have no influence on this estimate. Ignoring the fine structure of $(n-1)d+ns$ states, $|U_{ds}(R_0)|$ should be equal to $|C_6|/R_0^6$, where $C_6$ is the $C$ coefficient for the $(n\!-\!1)d+ns$ state.  The estimate for the lower limit $R_0$ is $R_0^{-3}\sim\sqrt{\Gamma/|C_6|}$. Since $C_6\sim n^{11}$, we finally find
\begin{equation} \label{n-scaling}
P_{{\rm exc}}\sim n^{8.5}.
  \end{equation}

\section{\label{sec:Results}Results and discussion}
\subsection{Numerical evaluation of excitation probabilities}
We have used four different ways to evaluate excitation probabilities in order to verify that the approximations are applicable for all conditions under which they are used. These conditions are different for different asymptotes. One way is to use the method \cite{Stanojevic} (based on Eq. (\ref{probab11})) to get the lineshape of $(n-1)d+ns$ resonances. However, it is assumed in this method that the effective two-photon Rabi frequency is sufficiently small and the molecular potentials are well separated. Therefore, for any pair of atoms at a certain separation $R$, only one doubly-excited molecular state is involved in the excitation (not necessarily the same one for all $R$). However, the two-photon Rabi frequency is almost an order of magnitude higher in this case so it is not obvious that some power-dependent terms can always be ignored. Such terms are included in Eq. (\ref{c_2Solut}).  We find that, for most potentials, the formulae (\ref{c_2Solut})-(\ref{probab11}) give overall very similar lineshapes, with the position of the resonance slightly shifted, but the shape and amplitude are well preserved. However, it turns out that the approximate Eqs. (\ref{probab11}) and (\ref{c_2Solut}), for the experimental parameters, cannot be used for all laser frequencies and  all asymptotes. By varying the laser frequency, we actually vary the region of internuclear separations $R$ for which atom pairs are on resonance. The two-photon Rabi frequency  $\omega_{\rm eff}$  is $R$-dependent and, if the $\ell$-mixing is too large, the approximations may not be valid. To check if such parameters significantly affect the final probabilities, we have numerically solved the system (\ref{Eqs-system-1})-(\ref{Eqs-system-6}). 

It is solved for many different internuclear separations, spatial orientations of the molecular axis, laser intensities and laser frequencies. This calculation is repeated for the states $\left|\varphi_{\lambda}\right>$  which have dominant contributions to the resonance lineshape.  They coincide with the asymptotic $69d+70s$,  $68d+71s$ and $68p_{1/2}+72p_{1/2}$ states. In these numerical calculations, only one molecular state $\left|\varphi_{\lambda}\right>$  is considered as a final doubly-excited state of an atom pair. As explained previously, at some $R$, one should consider superpositions of molecular states $|\varphi_{\lambda}\rangle$. This is taken into account in Eqs. (\ref{solution_tot}) and (\ref{truncate_solution}). Even though we do not consider the actual pulse shape in this case, the parameters of the substituted pulse are chosen to minimize any quantitative difference between the pulses. Since we use exact solutions of (\ref{hamilton_tot}), this approach should give a rather fair description of the resonance phenomenon.
\begin{figure}[t]
    \centerline{\epsfxsize=3.4in\epsfclipon\epsfbox{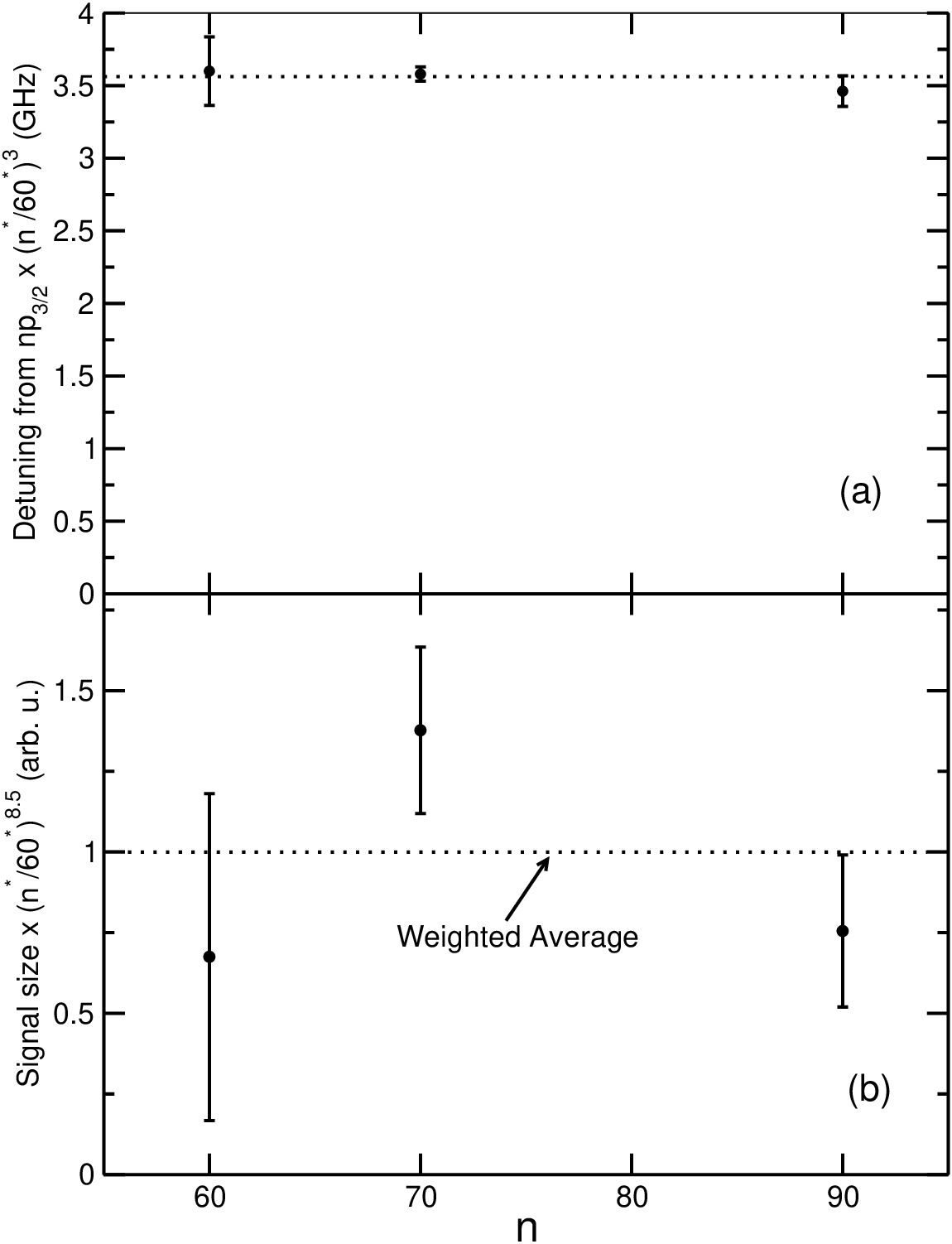}}
\caption{\label{scaling} (a) Scaling of the position of molecular $(n-1)d+ns$ resonances with respect to the atomic $np_{3/2}$ resonance. The position follows the characteristic $n^{-3}$ scaling. (b) Scaling of the molecular signal. Here we test the expected $n^{8.5}$ scaling (\ref{n-scaling}). This plot shows a reasonable agreement with the  $n^{8.5}$ scaling.}
\end{figure}

It turns out that all these different methods lead to the same physical results, although, for some particular parameters, they may be significantly different. Most of the differences vanish after summing over all potential curves and all atom pairs. The remaining variations are mainly due to the slight difference in the resonance positions and lineshapes obtained using different methods. The real physical parameters, such as the linewidth, signal size, $n$-scaling or even the resonance position, are practically unchanged, as shown in Fig. \ref{methods}. Here we present the result based on Eqs. (\ref{solution_tot})-(\ref{truncate_solution}), the most complete method we use, and the results from the simplest method based on Eq. (\ref{probab11}). This gives an estimate of the variations between these methods.
Our calculation shows that the probabilities of the $0_u^{-}$ asymptotes are about one order of magnitude less then the contribution of the other two symmetries. 
Interestingly, for $(n-1)p+(n+1)p$ resonances, the  contribution of the $0_g^{+}$ symmetry was insignificant \cite{Stanojevic}.

\subsection{Comparison with theoretical lineshapes}
In Figs. \ref{0Gsymm}(a)-\ref{1Usymm}(a), we show three sets of molecular potentials corresponding to the three symmetries  considered. At short distances these potentials have  very complicated shapes due to multiple avoided crossings. At these avoided crossings the $\ell$-mixing is the strongest. The $np\!+\!np$ components of various nearby potentials are the most important for the molecular resonance. The relevant physical quantity which depends on the fraction of $np\!+\!np$ states is the two-photon Rabi frequency $\omega_{\rm eff}$, defined by Eq. (\ref{ef-Rabi}). Asymptotes $(n\!-\!1)d+ns$, $(n\!-\!2)d+(n\!+\!1)s$ and $(n\!-\!2)p_{1/2}+(n\!+\!2)p_{1/2}$ give the essential contributions to the resonance.  In parts (b) of Figs. \ref{0Gsymm}-\ref{1Usymm} we present the magnitude of the radial dependence $\omega_{\rm eff}$ for the three asymptotes and all their states. For each of these states, the radial dependence is obtained after  averaging $\omega_{\rm eff}^2$ over different orientations of the molecular axis and initial states. Even though $\omega_{\rm eff}$ for atom pairs at short distances is larger due to stronger $\ell$-mixing, such pairs are difficult to excite because they are very detuned from the molecular resonance. The actual pair excitation probability as a function of $R$, on exact molecular resonance, is shown in Fig. \ref{radialProb}. The weighting factor $R^2$ is included in the presented dependence. Surprisingly, there is a significant contribution to the molecular signal from the pairs at larger distances.

We illustrate different methods used to calculate the resonance lineshape in Fig. \ref{methods}(b). We  present the simplest method, given by  Eq. (\ref{probab11}), and the method based on the exact solution for square pulses
(\ref{solution_tot}-\ref{truncate_solution}). Only the latter method allows superpositions of molecular states to be excited. This possibility is relevant if potentials are very close to each other. These two methods give overall very similar lineshapes, and the difference is only in the details.  The relative difference between them is somewhat larger just above the resonance, where there are many potentials very close to each other.  For the position of the molecular resonance, the method method based on Eqs. (\ref{solution_tot}-\ref{truncate_solution}) gives $-2.18$ GHz from the atomic $70p_{3/2}$ resonance, which is in agreement with the experimental position of 2.21(3) GHz. It appears that the position of the resonance obtained using the simplest method is closer to the experimental value. However, this is likely just a coincidence because, in that calculation $\Delta$ and $\Delta'$ were replaced by their average value, so that fine details of the atomic level positions were not included. 

We have tested the $n$-scaling of the molecular resonance, both numerically and experimentally. As mentioned, this scaling law makes sense only if, for a given laser intensity, the two-photon transition is not saturated for all values of $n$ for which the scaling law is used. The calculated ratio (using Eq. (\ref{probab11})) of the signals  for all $n=70$ and $n=60$ is 3.915, while $(70^{*}/60^{*})^{8.5}=3.920$. Also the same ratio for $n=90$ and $n=70$ is 9.274, while $(90^{*}/70^{*})^{8.5}=9.115$. The experimental dependence for the  $n$-scaling of the molecular signal is shown in Fig. \ref{scaling}(b). The agreement is fairly good and the deviation could be explained by variations of experimental parameters between these three experimental scans. The $n$-scaling of the resonance position is in an excellent agreement with the expected scaling law, as shown in Fig. \ref{scaling}(a).

\begin{figure}[t]
    \centerline{\epsfxsize=3.4in\epsfclipon\epsfbox{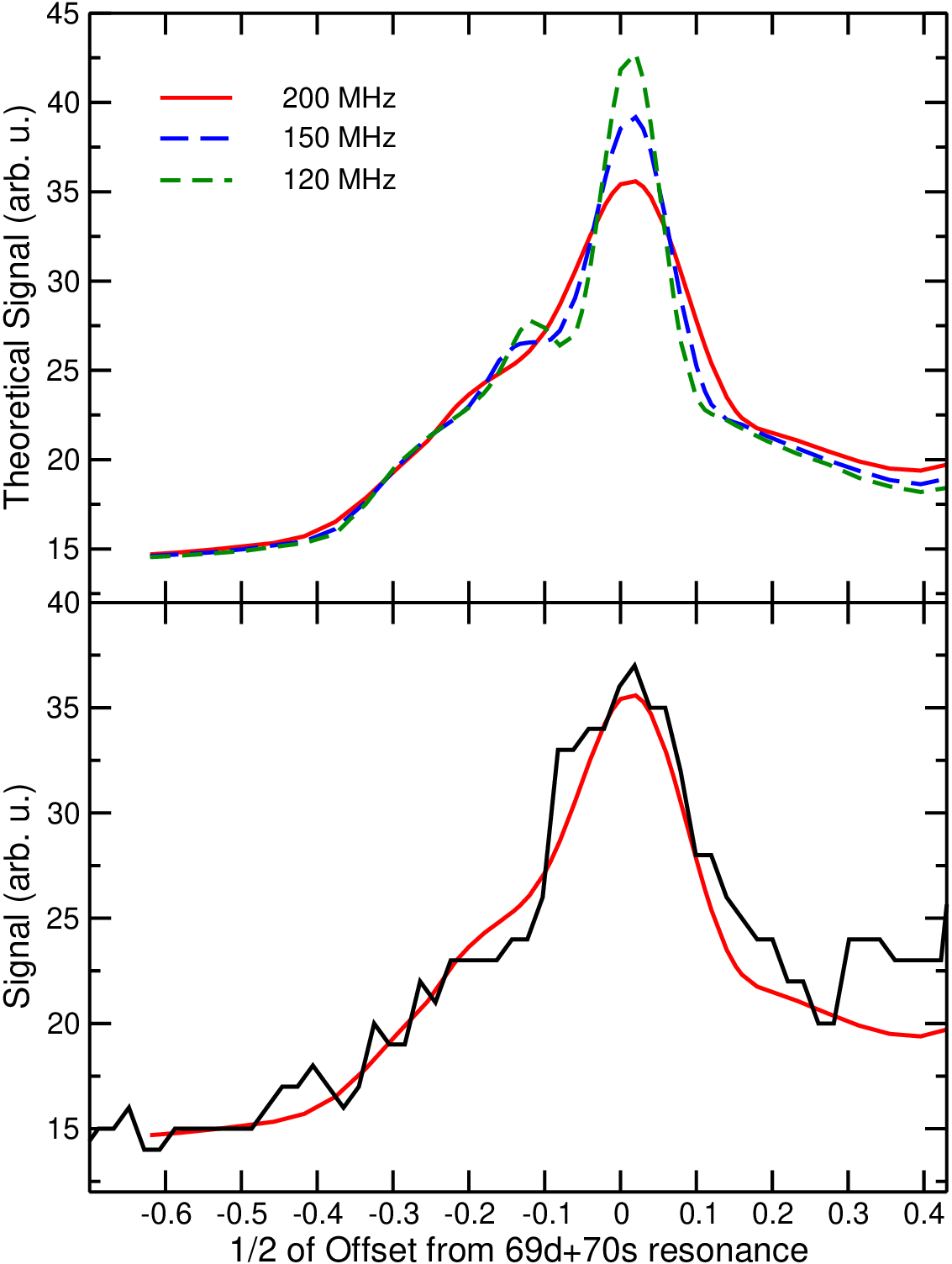}}
\caption{\label{convolut}(a) Calculated molecular signal for different laser bandwidths of the excitation laser, as indicated in the plot. The linewidth of the resonance is dominantly determined by the details of long-range interactions and it is significantly larger than the laser bandwidth. (b) Comparison of the  theoretical signals shown in Fig. \ref{methods}(b) and the experimental one. This is the best fit of the experimental data, although the actual bandwidth was likely smaller than the one we used for the comparison.}
\end{figure}

In Fig. \ref{convolut}(a) we present calculated lineshapes for several laser bandwidths, as indicated in the graph. The linewidth of the resonance is significantly larger than the laser bandwidth and is primarily determined by the details of the molecular potentials and $\ell$-mixing. The convoluted lineshapes for a laser bandwidth of 200 MHz is shown in Fig. \ref{convolut}(b). We get the best agreement with the experiment for this bandwidth. The actual laser bandwidth was probably smaller than this one. For the actual experimental conditions, this theory cannot be used to fit the portion of the red tail of the spectrum closer to the atomic resonance because the excitation fractions are much larger. The presence of excited atoms could modify the pair excitation probability so that Eq. (\ref{probab-av}) is not applicable.  Also, the simple pair excitation model cannot be used to explain the excitation process at larger excitation fractions. At such fractions, $\ell$-mixing could be more efficient because close molecular potentials could additionally mix due to the interactions with nearby excited atoms. This could explain why the experimental red tail, for the experimental conditions, is just a simple monotonic function even though there are many potential curves and avoided-crossings between the $np+np$ and $(n-1)d+ns$ asymptotic levels.

We compare experimental and theoretical signal sizes assuming a typical $5\times10^{10}$ cm$^{-3}$ density. The laser intensity is typically in the range of 400-500 MW cm$^{-2}$. Experimental signals are about 300 ions per shot. The  details of the experimental setup are given in \cite{farooqi03}. From the scaling factor introduced in order to compare the theoretical and experimental lineshapes, we find that the calculated number of atoms is about 6-7 times higher than the experimental value. This is probably acceptable considering the number of factors which influence this estimate. Besides uncertainties  in the experimental parameters, this scaling factor may also suggest that there was some excitation blockade \cite{tong04,singer04}.

\section{\label{sec:conclusion}Conclusion}
We have presented long-range doubly-excited molecular potentials of the $0_g^{+}$, $0_u^{-}$ and $1_u$ symmetries. These potentials are important in describing the effects of interactions in single-photon excitation to high $np$ Rydberg states. We have illustrated the $\ell$-mixing induced by interactions over a broad range of internuclear distances. Several methods to evaluate the spectrum of the excitation of interacting Rydberg pairs of atoms were presented and they all gave similar results. The analysis showed that  molecular resonances at the average atomic $nd$ and $ns$ energies are expected to occur. The calculated properties of these resonances, such as position, linewidth, $n$-scaling and signal size are reasonably close to the experimental observation, but a complete understanding of the spectral features of Rydberg excitation requires improved theoretical models. 

\begin{acknowledgments}
This work was supported in part by the National Science Foundation.
\end{acknowledgments}


\section*{References}

\begin{thebibliography}{99}
\bibitem{gallagher-book}
   T. F. Gallagher, Rydberg Atoms (Cambridge University Press,
   Cambridge, 1994).
\bibitem{oliveira03}
   A. L. Oliveira, M. W. Mancini, V. S. Bagnato, and
   L.G. Marcassa, Phys. Rev. Lett. {\bf 30}, 143002 (2003).
\bibitem{raimond81} 
   J. M. Raimond, G. Vitrant, and S. Haroche, 
   J. Phys. B {\bf 14}, L655 (1981).
\bibitem{jaksch00} 
   D. Jaksch, J. I. Cirac, P. Zoller, S. L. Rolston,
   R. C{\^o}t{\'e}, and M. D. Lukin, Phys. Rev. Lett. {\bf 85},
   2208 (2000).
\bibitem{lukin01} 
   M. D. Lukin, M. Fleischhauer, R. C{\^o}t{\'e}, L. M. Duan,
   D. Jaksch, J. I. Cirac, and P. Zoller, Phys. Rev. Lett.
   {\bf 87}, 037901 (2001).
\bibitem{tong04} 
   D. Tong, S. M. Farooqi, J. Stanojevic, S. Krishnan,
   Y. P. Zhang, R. C{\^o}t{\'e}, E. E. Eyler, and P. L. Gould, 
   Phys. Rev. Lett. {\bf 93}, 063001 (2004).
\bibitem{singer04}
   K. Singer, M. Reetz-Lamour, T. Amthor, L. G. Marcassa, 
   and M. Weidem\"uller, Phys. Rev. Lett. {\bf 93}, 163001 (2004). 
\bibitem{Liebisch} 
 T. Cubel Liebisch, A. Reinhard, P. R. Berman, and G. Raithel, Phys. Rev. Lett. {\bf 95} 253002  (2006).
\bibitem{voght06}
   T. Vogt, M. Viteau, J. Zhao, A. Chotia, D. Comparat, and P. Pillet, Phys. Rev. Lett.  \textbf{97}, 083003 (2006).
\bibitem{greene00}
   C. H. Greene, A. S. Dickinson, and H. R. Sadeghpour, Phys. Rev. Lett.
   \textbf{85}, 2458 (2000).
\bibitem{granger01}
   B. E. Granger, E. L. Hamilton, and C. H. Greene, Phys. Rev.
   A \textbf{64}, 042508 (2001).
\bibitem{hamilton02}
   E. L. Hamilton, C. H. Greene, and H. R. Sadeghpour, J. Phys. B \textbf{35},
   L199 (2002).
\bibitem{chibisov02}
   M. I. Chibisov, A. A. Khuskivadze, and I. I. Fabrikant, 
   Phys. Rev. A \textbf{66}, 042709 (2002).
\bibitem{khuskivadze02}
   A. A. Khuskivadze, M. I. Chibisov, and I. I. Fabrikant, 
   J. Phys. B \textbf{35}, L193 (2002).
\bibitem{boisseau02} 
   C. Boisseau, I. Simbotin, and R. C\^ot\'e, Phys. Rev. Lett. {\bf 88} 133004 
   (2002).
\bibitem{farooqi03} 
   S. M. Farooqi, D. Tong, S. Krishnan, J. Stanojevic,
   Y. P. Zhang, J. R. Ensher, A. S. Estrin, C. Boisseau, R. C{\^o}t{\'e},
   E. E. Eyler, and P. L. Gould, Phys. Rev. Lett. {\bf 91}, 183002 (2003).
\bibitem{Stanojevic}
   J. Stanojevic, R. C{\^o}t{\'e}, D. Tong, S.M. Farooqi, E.E. Eyler 
 and P.L. Gould, Eur. Phys. J. D {\bf 40}, 3 (2006).
\bibitem{Brown} J. M. Brown and A. Carrington,
  \textit{Rotation Spectroscopy of Diatomic Molecules},  Cambridge University Press (Cambridge 2003). 
\bibitem{shaffer06}
   A. Schwettmann, J. Crawford, K. R. Overstreet, and J. P. Shaffer, Phys. Rev. A {\bf 74}, 020701(R) (2006).
\bibitem{buehler}  
   R. J. Buehler and J. O. Hirschfelder, Phys. Rev. \textbf{83}, 628 (1951).
\bibitem{marinescu97} 
   M. Marinescu, Phys. Rev. A {\bf 56}, 4764 (1997).
\bibitem{gallagher03}
   L. Wenhui, I. Mourachko1, M. W. Noel, and T. F. Gallagher, Phys. Rev. A {\bf 67}, 052502 (2003).
\bibitem{Shabanova} 
 L.N. Shabanova and A.N. Khlyustalov, Opt. Spectrosc. (USSR),  {\bf 56}, 128 (1994).
\bibitem{Tong}D.Tong \textit{et al.},  to be published. 
\end{thebibliography}
\end{document}